\documentclass[prl,aps,epsf,twocolumn]{revtex4}
\usepackage[english]{babel}
\usepackage{graphicx}
\usepackage{epsf}
\usepackage{pstricks}

\renewcommand{\in}{\raise -3pt\hbox{\scriptsize in}}
\newcommand{\out}{\raise -3pt\hbox{\scriptsize out}}

\begin{document}
\title{
\hfill CPHT-RR-001.0104 \\
\vskip.1in
Deep electroproduction of
exotic hybrid mesons}

\author{I.V.~Anikin$^{a,b}$,\, B.~Pire$^a$,\,
 L.~Szymanowski$^c$,\, O.V.~Teryaev$^{b,d}$,\, S.~Wallon$^e$}
\affiliation{$^a$CPHT de l'{\'E}cole Polytechnique,
             91128 Palaiseau Cedex, France\footnote{Unit{\'e} Mixte de
             Recherche du CNRS (UMR 7644)} \\
             $^b$Bogoliubov Laboratory of Theoretical Physics,
             JINR, 141980 Dubna, Russia \\
             $^c$Soltan Institute , Warsaw, Poland\\
             $^d$CPT-CNRS F-13288 Marseille Cedex 9, France
             \footnote {Unit{\'e} Propre de Recherche du
             CNRS (UPR 7061)} \\
             $^e$LPT,
             Universit{\'e} Paris-Sud, 91405, Orsay, France
             \footnote{Unit{\'e} Mixte de
             Recherche du CNRS (UMR  8627)}}

\begin{abstract}
\vskip.1in
\noindent
We evaluate the leading order amplitude for the deep exclusive electroproduction
of an exotic hybrid meson in the Bjorken regime. We show that, contrarily to naive expectation,
this amplitude factorizes at the  twist 2 level and thus scales like usual meson electroproduction
when the virtual photon and the hybrid meson are longitudinally polarized. Exotic hybrid mesons may thus be studied
in electroproduction experiments at JLAB, HERA (HERMES) or CERN (Compass).
\end{abstract}
\maketitle

\noindent
The study of hadrons outside the constituent quark models multiplets
is an interesting topic \cite{Close, Jaffe, Carlson, Bernard} which is of fundamental
importance to understand the dynamics of quark confinement.
Deep exclusive meson electroproduction (see, for instance \cite{goeke,DiehlPR})
is well
described in the framework of the collinear approximation where generalized
parton distributions (GPDs) \cite{Muller:1994fv} and distribution amplitudes
encode the nonperturbative parts of a factorized amplitude \cite{Collins:1997fb}.
In this letter we focus on the investigation of the electroproduction of
an isotriplet exotic meson with $J^{PC} = 1^{-+}$
which we will denote as $H$.
Candidates for such  states include $\pi_{1} (1400)$
\cite{RPP} which is mostly seen through its $\pi \eta $ decay.
Our derivation applies as well to exotic hybrid mesons with other quantum 
numbers. However, our analysis of the normalization of the amplitude
is only valid for the $1^{-\, +}$ state.   
 Theoretically
these states are subject of intense studies \cite{Close},
mostly through lattice simulations \cite{Bernard}.
A naive argument based on a constituent quark picture of the exotic hybrid meson
would lead to the expectation  that the amplitude has a vanishing twist 2 component and then
that the rate of such processes is suppressed at large $Q^2$ with
respect to usual meson electroproduction.
This is not  true since the quark-antiquark correlator on the light cone includes
a gluonic component due to gauge invariance and leads to a leading twist
hybrid light-cone distribution. We study such a correlator in detail
and calculate its contribution to the  hybrid electroproduction amplitude.
The reactions $\gamma^* p\to H^0 p, 
\gamma^* p\to H^+ n, \gamma^* n\to H^0 n, \gamma^* n\to H^- p$,
or the coherent reaction on deuteron \cite{Cano} $\gamma^*
d \to H^0 d$  may be studied to experimentally access this
amplitude. In this letter we restrict ourselves
to the first of them  and to  longitudinally polarized virtual photon and hybrid meson.

\noindent
Let us consider the hybrid--to--vacuum matrix element of
the bilocal quark operators.
As mentioned before,
we suppose that the hybrid is the isotriplet state with
$J^{PC}=1^{-+}$ quantum numbers.
In the quark model where  mesons are described as quark-antiquark states, such
quantum numbers are forbidden.

\noindent
Non-local quark operators necessarily involve gluon operators
due to color gauge invariance.
The key problem is whether this gluon admixture
allows this quark matrix element to have
exotic quantum numbers such as $J^{PC}=1^{-+}$.
To answer  this question we define, as usual, the meson distribution amplitude through
 the Fourier transformed correlator taken at $z^2=0$,
\begin{eqnarray}
&&\langle H(p,\lambda)| \bar \psi(-z/2)\gamma_\mu [-z/2;z/2]
\psi(z/2) |0 \rangle =
\nonumber\\
&& if_H M_H \biggl[
\biggl( e_\mu^{(\lambda)}-p_{\mu}\frac{e^{(\lambda)}\cdot z}{p\cdot z} \biggl)
\int\limits_0^1 dy e^{i(\bar y - y)p\cdot z/2} \phi^{H}_T(y)
\biggr.
\nonumber\\
\biggl.
&&+ p_\mu \frac{e^{(\lambda)}\cdot z}{p\cdot z}
\int\limits_0^1 dy e^{i(\bar y - y)p\cdot z/2}\phi^{H}_{L}(y)\biggr]\;,
\label{hme}
\end{eqnarray}
where $\bar y = 1-y$;
$f_H$ denotes a dimensionfull coupling constant of the hybrid meson, so
that the distribution amplitude $\phi^H$ is dimensionless. We will discuss
its normalization later.

\noindent For the longitudinal polarization case
$$e^{(0)}_\mu = \frac{e^{(0)}\cdot z}{p\cdot z} p_\mu$$
only $\phi^{H}_{L}$ contributes, so that
\begin{eqnarray}
&&\langle H(p,0)| \bar \psi(-z/2)\gamma_\mu [-z/2;z/2]
\psi(z/2) | 0 \rangle =
\nonumber\\
&&i f_H M_H e^{(0)}_\mu
\int\limits_0^1 dy e^{i(\bar y - y)p\cdot z/2} \phi^{H}_L(y)\;.
\label{hmeW}
\end{eqnarray}
In Eqs. (\ref{hme}) and (\ref{hmeW}),
we insert the path-ordered gluonic exponential along the straight line connecting
the initial and final points $[z_1;z_2]$ which provides the gauge invariance
for bilocal operator and
equals unity in a light-like (axial) gauge. For  
simplicity of notation we shall omit from now on the index
$L$ from the  hybrid meson distribution amplitude.

\noindent
Let us now prove that it is possible to describe in this way an exotic  $J=1$ meson state with
quantum numbers  $PC = -+$.
From  charge conjugation invariance, one can immediately deduce for the neutral member
of the isotriplet $H^0$ with the flavour structure $1/\sqrt{2}(\bar u
u -\bar d d)$ that
the parameterizing function $\phi^{H}$ is
antisymmetric , {\it i.e.}
\begin{eqnarray}
\phi^{H}(y)=-\phi^{H}(1-y).
\label{asy}
\end{eqnarray}
Isospin invariance and G-parity imply the same relation for charged
hybrids. The property (\ref{asy}) is
similar to the case of two pion 
distribution amplitude \cite{DGPT}.
In particular, the antisymmetric property implies
\begin{eqnarray}
\int\limits_{0}^{1}dy \; \phi^{H}(y)=0.
\label{asy1}
\end{eqnarray}

\noindent
Let us pass to the analysis of the remaining quantum number, parity.
For this purpose, it is convenient to expand
the left hand side of (\ref{hme})  in a Taylor series,
which is possible because the matrix element is assumed to be UV
regularized and does not contain any 
singularities in $z$ (these singularities only appear
in the hard scattering coefficient function ).
As a result,
the hybrid--to--vacuum matrix element of this operator may be
rewritten in the form:
\begin{eqnarray}
\label{locdec}
&&\langle H(p,\lambda)| \bar\psi(-z/2) \gamma_{\mu}[-z/2;z/2] \psi(z/2)| 0\rangle=
\\
&&\sum_{n\, odd}\frac{1}{n!}z_{\mu_1}..z_{\mu_n} \langle H(p,\lambda)|
\bar\psi(0) \gamma_{\mu}
\stackrel{\leftrightarrow}{D}_{\mu_1}..\stackrel{\leftrightarrow}{D}_{\mu_n}
\psi(0)| 0\rangle ,
\nonumber
\end{eqnarray}
where $D_{\mu}$ is the usual covariant derivative and
$\stackrel{\leftrightarrow} {D_{\mu}}=\frac{1}{2}(
\stackrel{\rightarrow}{D_{\mu}}-
\stackrel{\leftarrow}{D_{\mu}}$).
Due to the positive charge parity of $H^0$, see (\ref{asy}), only
odd terms in (\ref{locdec}) do contribute.
The simplest case is provided by the $n=1$ twist 2 operator
\begin{eqnarray}
{\cal R}_{\mu\nu}=\mbox{\Large S}_{(\mu \nu)}
\bar\psi(0)\gamma_{\mu}
\stackrel{\leftrightarrow}{D}_{\nu}\psi(0),
\end{eqnarray}
where ${\Large S}_{(\mu\nu)}$ denotes the standard symmetrization
operator (${\Large S}_{(\mu \nu)}T_{\mu \nu}=1/2(T_{\mu \nu}+T_{\nu
\mu})$).
${\cal R}_{\mu \nu}$ is
proportional to the quark energy-momentum tensor, {\it i.e.}
 ${\cal R}_{\mu\nu} = - i \Theta_{\mu\nu}$.
Its matrix element of interest is
\begin{eqnarray}
\label{emee}
&&\langle H(p,\lambda) | {\cal R}_{\mu\nu} |0\rangle=
\nonumber\\
&& \frac{1}{2}\,f_H M_H \mbox{\Large S}_{(\mu \nu)}
e_{\mu}^{(\lambda)} p_{\nu} \int\limits_{0}^{1}dy (1-2y) \phi^{H}(y),
\end{eqnarray}
Note that it is the symmetry in $\mu \nu$ of the energy momentum tensor
which selects the twist-2 function.

\noindent
To determine the parity one should treat the meson polarization
with some care.
The equation $e_{L\,\mu} \sim p_{\mu}/M_H $ holds for a {\it fast}
longitudinally polarized vector meson. On the other hand,  the meson
is an eigenstate of the parity operator $P$ only in its {\it rest} frame.
In this frame $p_{\mu}$ has only a zeroth component, while $e_{\mu}$
has a vanishing zeroth component.
This
leads to the negative parity of the relevant components ${\cal R}_{0k}$ with
$k=1,2,3$.
\begin{eqnarray}
P\biggl( \mbox{\Large S}_{(k 0)}\bar\psi(0)\gamma_{k}
\stackrel{\leftrightarrow}{D}_{0}\psi(0)\biggr)= - .
\end{eqnarray}
This  explicitly shows that the non-local matrix element (\ref{hmeW}) may describe an exotic
hybrid meson and its light-cone distribution amplitude is a leading twist quantity
with vanishing first moment (\ref{asy}).

\noindent
The non-zero matrix element of the quark energy-momentum tensor between vacuum and
exotic meson state was explored long ago \cite{Koles}.
It may be related, by making use
of the equations of motion, to the matrix element of quark-gluon operator
and  estimated with the help of the techniques of QCD sum rules \cite{Balitsky},
which allows to fix the normalization factor, or the coupling constant, $f_{H}$.
One of the solutions corresponds to a resonance with  mass
around $1.4\, {\rm GeV}$ and the coupling constant at this scale
 \cite{our}
\begin{eqnarray}
\label{fh}
 f_{H } \approx 50 \,{\rm MeV}\;.
\end{eqnarray}

\noindent
Note  that the same exotic quantum numbers (except isospin) were found \cite{Jaffe}
for the gluonic energy momentum tensor (attributed therefore to
the gluonium). It was noticed there that  energy momentum conservation
leads to a zero coupling of the operator to such an exotic state.
This argument would be applicable in our case for the  isosinglet
combination, if
the quark gluon interaction, leading to the non-conservation of both
quark and gluon energy momentum tensor (while the sum is conserved),
is assumed to be negligible.
However, there is no reason to expect it to be applicable to isovector
combinations or to each quark flavour separately.
Moreover, even for isosinglet combination (including the pure gluonium case),
this argument is no more applicable to the
local operators of higher spin ($n=3,5...$).
The appearance of extra covariant derivatives:
\begin{eqnarray}
\label{eme}
&& \langle H(p,0) |{\cal R}_{\mu \nu_1...\nu_n} |0\rangle=
\nonumber\\
&& i^{n+1} f_H M_{H} \mbox{\Large S}_{(\mu \nu_1 ... \nu_n)}
e^{(0)}_\mu  p_{\nu_1}... p_{\nu_n}
\nonumber\\
&& \times \,\int\limits_{0}^{1}dy (y-\frac{1}{2})^n
\phi^{H}(y),
\end{eqnarray}
preserves all the quantum numbers, but spoils the argument of the
operator conservation, as it is not the energy-momentum tensor anymore.
Such a situation is completely similar to the case of tensor
spin structure or fragmentation function \cite{ET82,SST99},
which have two zero moments.

\noindent
In summary, the hybrid light-cone distribution amplitude is a leading twist quantity
which should have
a vanishing first moment (\ref{asy})  because
of the antisymmetry.
This distribution amplitude obeys usual evolution equations \cite{ERBL}
and has an asymptotic limit \cite{Chase}
\begin{eqnarray}
\Phi^H_{as} = 30  y (1-y) (1-2y)
\end{eqnarray}
with assumed normalization of the distribution amplitude $\phi^H(y)$ as
\begin{equation}
\label{normal}
\int\limits_0^1 dy (1-2y)\phi^H(y)=1\;.
\end{equation}
The coupling constant $f_H$ is the subject of evolution given by the
formula, see e.g. \cite{DGP}
\begin{equation}
\label{evolution}
f_H(Q^2)= f_H\cdot \left(\frac{\alpha_S(Q^2)}{\alpha_S(M_H^2)} 
\right)^{K_1}\;\;\;\;K_1=\frac{2\,\gamma_{QQ}(1)}{\beta_1}\;,
\end{equation}
where the anomalous dimension $\gamma_{QQ}(1)=16/9$ and
$\beta_1=11-2n_f/3$.
The exponent $K_1$ is thus 
 a small positive number which drives slowly to zero the coupling
 constant $f_H(Q^2)$. Since experiments are likely to be feasible
 at moderate values of $Q^2$, we neglect this evolution and in the
 following estimate we use the value from Eq. (\ref{fh}).


\noindent
The calculation of the production amplitude, at leading order in
$\alpha_s$, is now straightforward
and leads to an expression completely similar to the
one for the production of longitudinally polarized vector 
meson, see e.g. \cite{goeke} and notation therein. It is well known that 
a leading twist estimate of the $\rho$ electroproduction cross-section 
underestimates the experimental rate, but we 
think that it is still reasonable to estimate the ratio 
of hybrid to $\rho$ electroproduction cross-sections 
through their Born order expression. We obtain:
\begin{eqnarray}
\label{qdsim2}
{\cal A}_{\gamma^*_L p\to H^0_L p}=&&  \frac{e\pi\alpha_s
f_{H}\,C_F}{\sqrt{2}\;N_c\,Q}
\nonumber\\
&& \times \biggl[ e_u {\cal H}_{uu}^- -e_d {\cal H}_{dd}^-\biggr] {\cal V}^{H},
\end{eqnarray}
where
\begin{eqnarray}
\label{softin}
{\cal H}_{ff}^\pm=
\int\limits_{-1}^{1}dx &&\biggl[
\overline{U}(p_2) \hat{n} U(p_1) H_{ff}(x,\xi) +
\biggr.
\nonumber\\
\biggl.
&&\overline{U}(p_2)\frac{i\sigma_{\mu\alpha}
n^{\mu}\Delta^{\alpha}}{2M}U(p_1)
E_{ff}(x,\xi) \biggr]
\nonumber \\
&&
\times \, \biggl[
\frac{1}{x+\xi-i\epsilon} \pm \frac{1}{x-\xi+i\epsilon}\biggr],
\end{eqnarray}
and
\begin{eqnarray}
{\cal V}^{H} =
\int\limits_{0}^{1} dy \phi^{H}(y)\biggl[
\frac{1}{y}-\frac{1}{1-y}
\biggr].
\nonumber
\end{eqnarray}
Note that the simple pole in $y$ in (\ref{qdsim2}) does not lead to
any infrared divergency since the function $\phi^{H}(y)$ is expected to vanish,
as usual, when the fraction $y$ goes to zero or unity.

\noindent
We leave for a further work  a full phenomenological study of hybrid electroproduction.
Let us just note that the order of magnitude of hybrid electroproduction may be easily
deduced through a direct comparison with $\rho$ meson electroproduction
amplitude \cite{goeke}. We
thus estimate that the ratio of hybrid and $\rho$ 
electroproduction
cross-sections is:
\begin{eqnarray}
\label{ratio}
\frac{d\sigma^{H}(Q^2, x_B, t )}{d\sigma^{\rho}(Q^2, x_B, t )}=
\biggl|\frac{f_H}{f_\rho}
\frac{( e_u {\cal H}_{uu}^- -e_d {\cal H}_{dd}^-) {\cal V}^{H}}
{( e_u {\cal H}_{uu}^+ -e_d {\cal H}_{dd}^+) {\cal V}^{\rho}}\biggr|^2,
\nonumber \\
\end{eqnarray}
where the $\rho$ meson soft integral ${\cal V}^{\rho}$ is defined as
\begin{eqnarray}
{\cal V}^{\rho} =
\int\limits_{0}^{1} dy \phi^{\rho}(y)\biggl[
\frac{1}{y}+\frac{1}{1-y}
\biggr],
\nonumber
\end{eqnarray}
and $f_\rho = 216\,$MeV. 
Note that the symmetry under $x\to -x$  of the
nucleon GPD's in the numerator of (\ref{ratio}) corresponds to the
"non-singlet", {\it i.e.} $q-\bar q$, combination of quark distributions,
while the antisymmetric form in the denominator
 -- the "singlet", {\it i.e.} $q+\bar q$, combination.\\
If we neglect the antiquarks contribution, {\it i.e.} if we restrict
the $x$-integral to $[0,1]$, one sees  that the imaginary parts of the
amplitudes for both meson electroproduction are equal in magnitude up
to the factor ${\cal V}^{M}$. The ratio of the real parts depend
much on the model used for guessing the generalized parton
distributions. Since the imaginary part dominates in some kinematics, it is not unreasonable as
a first estimate of the ratio of the cross sections, to assume that the
full amplitude ratio is driven by the same quantity.
Using the
asymptotic forms for the hybrid and $\rho$ mesons
distribution amplitudes, which for the $\rho-$meson case is supported
by QCD sum rule \cite{qcdsr},  we thus estimate that :
\begin{eqnarray}
\frac{d\sigma^{H}(Q^2, x_B,  t )}{d\sigma^{\rho}(Q^2, x_B, t )}
\approx \biggl( \frac{5 f_H}{3 f_\rho}\biggr)^2 \approx 0.15.
\end{eqnarray}
Exotic hybrid meson can be therefore electroproduced in an
experimentally  feasible way in actual experiments at JLAB, HERMES or
Compass. Their study in high statistics experiments at JLAB should be
fruitful.
The signal may be discovered through a missing mass
measurement provided the recoil proton energy-momentum is well measured. This allows
to study all decay channels of these poorly known states. If one decay channel turns out to be
dominant, as for instance a $\pi \eta$ channel \cite{Adams},
the formalism of generalized distribution
amplitudes \cite{GDA2} may be used for estimating cross sections and interference
signals \cite{PO}.

\noindent
As already noted above, 
  higher twist 
corrections are likely to be sizable 
at measurable $Q^2$ \cite{Vand99}.
Twist 3 contributions have already been
considered in the case of deeply virtual Compton scattering \cite{APT} where their presence
was dictated by gauge invariance. They have also been considered for transversely polarized
vector mesons
\cite{AT} where the leading twist 2 component vanishes. The analysis of such
contributions  is left for future work.

\noindent
Let us finally note that  hard exclusive electroproduction turns out
to be a useful tool not only for studying the hybrid meson discussed 
in this paper but also for probing the structure of
other exotic states like the recently discovered pentaquark \cite{DPS}.
In all cases, the scaling of the amplitudes is the same as the observed 
one for $\rho$ electroproduction. The normalization of the cross sections is a more delicate strory, and may turn in some cases to unobservable rates 
\cite{ROPER}.

\noindent
We acknowledge useful discussions with A.P.~Bakulev, I.~Balitsky, V.~Braun,
M.~Diehl, G.~Korchemsky and  C.~Roiesnel.
This work is supported in part by INTAS (Project 00/587) and RFBR (Grant 03-02-16816).
The work of B. P. , L. Sz. and S. W. is partially
supported by the French-Polish scientific agreement Polonium.


\begin{thebibliography}{99}
\bibitem{our}
Our $f_{H}$
corresponds to $2 \sqrt{2} f_{R}$ in the notations of Ref. 
\cite{Balitsky}.
\end{thebibliography}
\end{document}